\documentstyle [aps,twocolumn,epsfig]{revtex}
\catcode`\@=11
\newcommand{\be}{\begin{equation}}
\newcommand{\ee}{\end{equation}}
\newcommand{\bea}{\begin{eqnarray}}
\newcommand{\eea}{\end{eqnarray}}

\newcommand{\up}{\uparrow}
\newcommand{\down}{\downarrow}

\newcommand{\ri}{\mbox{i}}

\begin {document}

\title{ Theory Of The One Dimensional SU(4) Spin-Orbital Model}
\vspace{2cm}

\author{P. Azaria$^1$,    A. O. Gogolin$^2$, P. Lecheminant$^3$,
  and A. A. Nersesyan$^{4,5}$}

\vspace{0.5cm}

\address{$^1$ Laboratoire de Physique Th\'eorique des Liquides,
Universit\'e Pierre et Marie Curie, 4 Place Jussieu, 75252 Paris,
France\\
$^2$ Departement of Mathematics, Imperial College, 180 Queen's Gate,
 London SW72BZ, United Kingdom\\
$^3$ Laboratoire de Physique Th\'eorique et Mod\'elisation,
Universit\'e de Cergy-Pontoise, Site de Saint Martin,
2 avenue Adolphe Chauvin, 95302
Cergy-Pontoise Cedex, France\\
$^4$ The Abdus Salam International Centre for Theoretical Physics, 
P.O.Box 586, 34100, 
Trieste, Italy \\
$^5$ Institute of Physics, Tamarashvili 6,
380077, Tbilisi, Georgia}
 
\vspace{3cm}

\address{\rm (Received: )}
\address{\mbox{ }}
\address{\parbox{14cm}{\rm \mbox{ }\mbox{ }
The one-dimensional  spin-orbital model is studied
by means of Abelian bosonization. 
We derive the low-energy effective theory which
enables us to study small
deviations from the SU(4) symmetric point. We show that 
there exists a massless region  with algebraically decaying
correlation functions, 
$\sim \cos(\frac{\pi}{2a_0} x) x^{-3/2}$.  When entering the massive phase, 
the system displays an approximate SO(6) enlarged 
symmetry with a dimerization type of ordering consisting in alternating 
spin and orbital singlets.
}}
\address{\mbox{ }}
\address{\parbox{14cm}{\rm PACS No: 75.10.Jm, 75.40.Gb}}
\maketitle

\makeatletter
\global\@specialpagefalse 
\makeatother
The interest in spin-orbital models stems from
the possibility of understanding the magnetic structures of transition 
metal compounds\cite{kugel}. 
In most of these materials, in addition to the usual spin degeneracy,
the low-lying electron states are also characterized by orbital degeneracy.
It is thus believed that the unusual magnetic properties observed in many  of
these compounds should be explained in terms of simple multi-band 
Hubbard-like models.
Very recently, the discovery of new spin-gapped
materials, Na$_2$Ti$_2$Sb$_2$O\cite{axtell} and Na$_2$V$_2$O$_5$\cite{isobe},
have attracted renewed interest in the spin-orbital models. 
These materials have a quasi-1D structure\cite{pati} and
are modelled 
by a quater-filled $two$-band Hubbard model which, in the limit of
strong Coulomb repulsion, is equivalent to two interacting Heisenberg models
with the Hamiltonian:
\bea 
{\cal H} = \sum_{i}   J_1 \; {\vec S}_{i} \cdot {\vec S}_{i+1}
&+& J_2 \; {\vec T}_{i} \cdot {\vec T}_{i+1} \nonumber \\
&+& K\left({\vec S}_{i} \cdot {\vec S}_{i+1}\right)
\left({\vec T}_{i} \cdot {\vec T}_{i+1}\right)
\label{hamilt}
\eea
where ${\vec S}_{i}$ and ${\vec T}_{i}$ are spin-1/2 operators
representing the spin and orbital degrees of freedom at each site $i$,
and $J_{1,2}$ and $K$ are positive constants.

The Hamiltonian (\ref{hamilt}) is invariant under independent SU(2) rotations
in the spin ($\vec S$) and orbital ($\vec T$) spaces.
It can also be recast 
as a two-leg spin ladder with a four-spin interchain coupling. 
In the limit $K \ll J_{1,2}$ this interaction, which can be generated either 
by phonons or (in the doped state) by the 
Coulomb repulsion between the holes,
gives rise to 
a non-Haldane spin-liquid state where magnon excitations are  
incoherent\cite{nersesyan}.
The physically relevant
question is whether or not this scenario can be 
extended to larger values of $K$ for
which (\ref{hamilt}) is expected to be of experimental relevance. 

As a matter of fact, we already
know that this cannot be the case. Indeed, the interesting feature 
of the Hamiltonian (\ref{hamilt})
is that at
$J_1 = J_2 =K/4$ it is not only SU(2)$\times$SU(2) symmetric
but actually has an enlarged SU(4) symmetry\cite{li}.
At this special point, the model is Bethe-ansatz solvable\cite{sutherland} and
critical with $three$ gapless bosonic modes; in  
the Conformal Field Theory language, that means that 
the central charge is $c=3$ and, as shown by Affleck\cite{affleck},
 the critical theory corresponds to 
the SU(4)$_1$ Wess-Zumino-Novikov-Witten (WZNW) model. 
Clearly, there should be a {\sl qualitative} change in the physical behavior
of (\ref{hamilt}) when going from small to large values of $K$. 
>From the theoretical point of view this situation is striking 
because it implies that one cannot go continuously from weak to strong 
coupling.
This is a manifestation of Zamolodchikov's c-theorem which states that,
starting at $K=0$ with two decoupled S=1/2 Heisenberg chains with the 
total central charge $c=2$
(two gapless bosons), one cannot flow - in the 
Renormalization Group (RG) sense -
towards the SU(4) point which has a larger central charge $c=3$.
Therefore, the physics in the neigborhood of the SU(4) point  
cannot be understood
in terms of weakly coupled 
Heisenberg chains, and the general strategy
employed to tackle spin ladders does not apply here: a new effective
theory is to be developed.
 It is the purpose of this work to do so.
Below we present
an  effective 
continuum description of the model
(\ref{hamilt}) at the SU(4) 
point, based on Abelian bosonization, and derive the low-energy expressions
for the spin and orbital densities. With these results at hand, we then 
investigate the properties of differing phases occuring at 
small deviations from 
the  SU(4) point.

-{\sl Abelian bosonization at the SU(4) point}.
We start by introducing the SU(4) Hubbard model with $U>0$:
\bea
{\cal H}_U &=& \sum_{ia\sigma} \left(-t \;c_{i+1 a\sigma}^{\dagger}
c_{ia\sigma} + \rm{H.c.}\right) \nonumber \\
&+ &
\frac{U}{2} \sum_{iab\sigma\sigma^{'}} n_{ia\sigma}  n_{ib\sigma^{'}}
\left(1 - \delta_{ab} \delta_{\sigma\sigma^{'}}\right).
\label{hubbard}
\eea
Here $c_{ia\sigma}^{\dagger}$ 
creates an electron
with the 
``flavor''(orbital) index $a=1,2$ and spin $\sigma = \up,\down$, 
and $n_{ia\sigma} = c_{ia\sigma}^{\dagger} c_{ia\sigma}$. 
It will be assumed that the electron band is
quarter-filled 
implying that the Fermi momentum 
$k_F = \pi/4a_0$, where $a_0$ is the lattice spacing.
The spin and orbital operators are defined as
\be
{\vec S}_{i} = \frac{1}{2} \sum_a c_{ia\alpha}^{\dagger} 
{\vec \sigma}_{\alpha\beta} c_{ia\beta},~~
{\vec T}_{i} = \frac{1}{2} \sum_{\alpha} c_{ia\alpha}^{\dagger} 
{\vec \tau}_{ab} c_{ib\alpha}
\label{spinrep}
\ee  
where ${\vec \sigma}$ (${\vec \tau}$)
are the Pauli matrices acting in the spin (orbital) space.

The low-energy physics can be described in terms of 
right-moving ($R_{a\sigma}$) and left-moving ($L_{a\sigma}$) 
fermions which replace the original
lattice fermion $c_{ia\alpha}$ in the continuum limit ($x=ia_0$):
\be 
\frac{c_{ia\sigma}}{\sqrt a_0} \simeq
R_{a\sigma}\left(x\right) \exp\left(\ri k_F x \right) + 
L_{a\sigma}\left(x\right) \exp\left(-\ri k_F x\right).
\label{ccont}
\ee
At this point we
introduce 
four chiral bosonic fields $\Phi_{a\sigma R,L}$ 
using Abelian bosonization of Dirac fermions:
$R(L)_{a\sigma}=\kappa_{a\sigma}(2\pi a_0)^{-1/2}
\exp\left(\pm\ri \sqrt{4\pi} \Phi_{a\sigma R(L)}\right)$.
The bosonic fields satisfy the 
commutation 
relation $\left[\Phi_{a\sigma R},\Phi_{b\sigma^{'} L}\right]=\frac{i}{4} 
\delta_{ab}\delta_{\sigma\sigma^{'}}$.
Anticommutation between the fermions with 
different spin-channel indices is
ensured by 
Klein factors (here Majorana fermions) $\kappa_{a\sigma}$.
It is then suitable to employ 
a physically transparent basis (cf. Ref. \cite{emery}):
\bea
\Phi_c &=& 
\left(\Phi_{1\up} + \Phi_{1\down} + 
\Phi_{2\up} + \Phi_{2\down} \right)/2 \nonumber \\ 
\Phi_s &=& 
\left(\Phi_{1\up} - \Phi_{1\down} +
\Phi_{2\up} - \Phi_{2\down}\right)/2  \nonumber \\ 
\Phi_f &=& 
\left(\Phi_{1\up} + \Phi_{1\down} -
\Phi_{2\up} - \Phi_{2\down}\right)/2   \nonumber \\ 
\Phi_{sf} &=& 
\left(\Phi_{1\up} - \Phi_{1\down} -
\Phi_{2\up} + \Phi_{2\down}\right)/2. 
\label{sfsf}
\eea
In the new basis, the total charge degree of freedom are described by 
$\Phi_c$, while the non-Abelian (spin-orbital) 
degrees of freedom,
are faithfully  represented by  
three bosonic fields $\Phi_a$ ($a = s,f,sf$).
It is now straightforward to obtain the continuum limit  of 
the Hubbard Hamiltonian (\ref{hubbard})
which exhibits separation between the charge and spin-orbital parts
of the spectrum.
The charge sector is described by a Gaussian model for the field
$\Phi_c$ perturbed by an  
Umklapp term $\sim \cos \sqrt{16 \pi K_c} \Phi_c$
generated in 
higher orders of perturbation theory. 
Though at small $U$ the Umklapp term 
is irrelevant and the charge excitations remain
gapless, one expects that 
on increasing the Coulomb interaction
the non-universal parameter $K_c(U)$ will 
decrease and eventually reach
the critical value $K_c (U_c)= 1/2$
where a Mott transition occurs to an insulating phase\cite{affleck,assaraf}. 
Though one certainly expects the system to be insulating in the limit 
$U/t \rightarrow \infty$\cite{kugel,auerbach,yamashita},
the question 
whether a commensurability gap, $m_c$, 
opens at a $finite$ value of $U$ is beyond the scope
of perturbation theory. 
Very recently, Assaraf et al\cite{assaraf} using an improved
Monte Carlo method were able to show that there 
exists a critical value $U = U_c \sim 2.8 t$ 
above which $m_c \neq 0$. Assuming
$U > U_c$, in what follows we shall focus on the spin-orbital sector
described by the Hamiltonian:
\bea
{\cal H}_{\em so} &=& \sum_{a=s,f,sf} \left[
\frac{v_F}{2} \left(\left(\partial_x \Phi_a\right)^2
+ \left(\partial_x \Theta_a\right)^2 \right)
+ \frac{G_3}{ \pi} \left(\partial_x \Phi_a\right)^2 \right]\nonumber \\
&-&  \frac{G_3}{\pi^2 a_0^2} 
\sum_{a \neq b} \cos\sqrt{4\pi}\Phi_a\cos\sqrt{4\pi}\Phi_b , 
\label{spinorbsepar}
\eea 
where $G_3 = - Ua_0/2$, and $\Theta_a =\Phi_{a L} - \Phi_{a R}$ are
the fields dual to $\Phi_a$.
The structure of the last term in (\ref{spinorbsepar}) immediately suggests
refermionization of the three bosonic fields $\Phi_a$ in terms of six real
(Majorana) fermions $\xi^a, a=(1,..,6)$:
\bea
\left(\xi^1 +\ri \xi^2\right)_{R(L)} &=& \frac{\eta_1}{\sqrt{\pi a_0}}
\exp\left(\pm \ri \sqrt{4\pi} \Phi_{sR(L)}\right) \nonumber \\
\left(\xi^3 + \ri \xi^4\right)_{R(L)} &=& \frac{\eta_2}{\sqrt{\pi a_0}}
\exp\left(\pm \ri \sqrt{4\pi} \Phi_{fR(L)}\right) \nonumber \\
\left(\xi^5 + \ri \xi^6\right)_{R(L)} &=& \frac{\eta_3}{\sqrt{\pi a_0}}
\exp\left(\pm\ri \sqrt{4\pi} \Phi_{sfR(L)}\right),
\label{majorep1}
\eea
$\eta_i$ being another Klein factors. In this representation the original
SU(4) transformations of the complex fermion fields appear as
SO(6) rotations
on the Majorana sextet $\{ \xi^a \}$, reflecting the 
equivalence SU(4) $\sim$ SO(6).
 To get a better insight in the symmetry properties of our model,
let us define the spin and orbital triplets:
$ {\vec \xi}_s = \left(\xi^2, \xi^1, \xi^6\right)$ and 
$ {\vec \xi}_t = \left(\xi^4, \xi^3, \xi^5\right)$.
Those transform as vectors under spin SO(3)$_s$ and
orbital SO(3)$_t$ rotations, respectively.
In the Majorana representation, the Hamiltonian  (\ref{spinorbsepar})
reduces to an SO(6) Gross-Neveu (GN) model: 
\be
{\cal H}_{so} = -\frac{iv_s}{2} \sum_{a=1}^{6} 
\left(\xi^{a}_R \partial_x \xi^{a}_R
- \xi^{a}_L \partial_x \xi^{a}_L\right) 
+ G_3 \left(\sum_{i=1}^{6} \kappa_i\right)^2
\label{grossneveu}
\ee
with $\kappa_i= \xi^{i}_{R} \xi^{i}_L$.
Since $G_3 <0$, we conclude that the interaction term in (\ref{grossneveu}) 
is marginally
irrelevant and 
the model flows towards six decoupled massless real fermions.
Thus, at the fixed point ($G_3^* =0$), the spin-orbital sector is described
by the SO(6)$_1$ ($\sim$ SU(4)$_1$) WZNW model with
the central charge $c = 6 \cdot 1/2 = 3$. 

To complete our description of the SU(4)-symmetric critical point,
we present
the continuum expressions
for the effective spin and orbital densities:
\bea 
{\vec S}& = &{\vec J}_{sR} + {\vec J}_{sL}  
+ \exp\left(\ri \pi x/2a_0\right) {\vec {\cal N}_s} + \rm{H.c.} + 
\left(-1\right)^{x/a_0} {\vec n}_s
 \nonumber \\
{\vec T} &=& {\vec J}_{tR} + {\vec J}_{tL}
+ \exp\left(\ri \pi x/2a_0\right) {\vec {\cal N}_t} + \rm{H.c.} 
+ \left(-1\right)^{x/a_0} {\vec n}_t
\label{spindensities2}
\eea
Here ${\vec J}_{s,t}$ are the smooth ($k\sim 0$) parts of these densities,
while ${\vec {\cal N}_{s,t}}$ and ${\vec n}_{s,t}$ are the 
$2k_F=\pi/2a_0$ and $4k_F=\pi/a_0$ parts.
Notice that Eqs. (\ref{spindensities2}) have a more complicated
structure than that of the spin density in the usual Hubbard model.
The emergence of 
$4a_0$-oscillations and
the corresponding complex fields ${\vec {\cal N}_{s,t}}$
is a consequence of the band's quarter-filling.
The smooth and $2k_F$ contributions can be computed directly from
Eqs. (\ref{spinrep}). We find that the chiral vector currents
$$
{\vec J}_{sR(L)} = - \frac{\ri}{2} \; {\vec \xi}_{sR(L)} 
\wedge {\vec \xi}_{sR(L)}, ~
{\vec J}_{tR(L)} = - \frac{\ri}{2} \; {\vec \xi}_{tR(L)} 
\wedge {\vec \xi}_{tR(L)}
$$
are in fact SU(2)$_2$ currents, in contrast with a single Heisenberg chain
where the smooth part of the spin density is a sum of
SU(2)$_1$ vector currents.
The fields ${\vec {\cal N}_{s,t}}$ are nonlocal in the Majorana fermions
$\vec{\xi_{s,t}}$. However, as in the 
two-leg ladder problem\cite{shelton}, they acquire a local form when 
expressed in terms
of order and disorder operators $\sigma_a$
and $\mu_a$ of the six critical Ising models associated with 
the six Majorana fermions. The expressions of 
${\vec {\cal N}_{s,t}}$ are 
manifestly SO(3)$_{s,t}$ invariant; here we give
only their $z$ components:
${\cal N}_{s}^z = 
A \left(\ri \mu_1 \mu_2 \sigma_3 \sigma_4 \sigma_5 \sigma_6
+ \sigma_1 \sigma_2 \mu_3 \mu_4 \mu_5 \mu_6 \right)$
and 
${\cal N}_{t}^z = 
A \left(\ri \sigma_1 \sigma_2 \mu_3 \mu_4 \sigma_5 \sigma_6
+ \mu_1 \mu_2 \sigma_3 \sigma_4 \mu_5 \mu_6 \right)$,
where $A$ is 
a non-universal constant. 
At the 
critical point, the order and disorder operators have scaling dimension $1/8$,
so the $2k_F$ densities ${\vec {\cal N}_{s,t}}$ have dimension
$3/4$. Since the $\vec{S}$ and $\vec{T}$ densities involve fermionic bilinears,
it may appear surprising to find 4$k_F$ contributions
${\vec{n}_{s,t}}$.
However, nothing prevents higher harmonics to be
generated in interacting systems. 
The structure of ${\vec{n}_{s,t}}$ can be 
anticipated
by symmetry arguments: these fields should be chirally invariant
and transform as vectors under SO(3)$_{s,t}$ rotations. These requirements lead
to the following simple expressions:
${\vec n}_s = \ri B \; {\vec \xi}_{sR} \wedge {\vec \xi}_{sL}$ and
${\vec n}_t = \ri B \; {\vec \xi}_{tR} \wedge {\vec \xi}_{tL}$, where $B$
is another non-universal constant. The scaling dimension 
of the ${\vec{n}_{s,t}}$ fields 
is 1.

-{\sl Deviations from the SU(4) point}.
We are now in a position to investigate the properties of the model
(\ref{hamilt}) at small deviations from the SU(4) point. 
We shall restrict consideration 
to symmetric perturbations, $J_1 = J_2 = K/4 + G$, $~|G|\ll K$,
and postpone
the study of a more general case
to a future publication.
Using the low-energy representation of the spin-orbital densities,
one can expand (\ref{hamilt}) around 
the SO(6)$_1$ fixed point to find:
\bea
{\cal H} = &-& iu/2 \left({\vec \xi}_{sR} \cdot \partial_x {\vec \xi}_{sR} -
{\vec \xi}_{sL} \cdot \partial_x {\vec \xi}_{sL} \right) \nonumber \\ 
&-& iu/2\left({\vec \xi}_{tR} \cdot \partial_x {\vec \xi}_{tR} -
{\vec \xi}_{tL} \cdot \partial_x {\vec \xi}_{tL} \right) \nonumber \\
&+&   G_3 \;\left( \kappa_1 + \kappa_2 + \kappa_3  + 
\kappa_4 + \kappa_5 + \kappa_6 \right)^2 \nonumber \\
&+& G \left[ \left( \kappa_1 + \kappa_2 + \kappa_6 \right)^2 +
 \left( \kappa_3 + \kappa_4 + \kappa_5 \right)^2 \right].
\label{hpert}
\eea
The Hamiltonian
(\ref{hpert}) describes two  SO(3)-symmetric, marginally coupled,
spin and orbital GN models. The $G$-term breaks SU(4)$\sim$ SO(6)
symmetry down to
SO(3)$_s \otimes$ SO(3)$_f$.
Notice that  all 
interactions are  marginal. This is the reason why we have also kept 
the marginally irrelevant ($G_3$) term which is already present 
at the SU(4) point (Eq. (\ref{grossneveu})). 
The emerging picture is to be opposed to
the case of two weakly coupled Heisenberg chains where the interchain
interaction $J_{\perp}$ gives rise to a strongly relevant perturbation
(of scaling dimension $1$) and thus opens a spectral gap
at arbitrarily small $J_{\perp}$\cite{nersesyan}. The RG
equations for the couplings in (\ref{hpert}) are easily 
obtained at the one-loop level:
\be
\dot{G} = G^2 - 2 G G_3, ~~
\dot{G}_3 = 4G_3 (G + G_3 ).
\label{rg}
\ee
The flow analysis reveals the existence of
three different regions: A, B and C, shown in Fig.1.
\begin{figure}
        \begin{center}
        \mbox{\psfig{figure=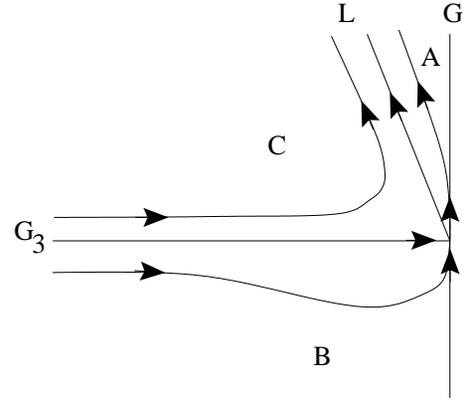,width=6cm}}
        \caption[99]{
        Flow diagram for isotropic couplings.}
        \end{center}
\end{figure}
In the region B, all couplings are 
irrelevant and a model with initial conditions
in B flows towards the SO(6)$_1$ fixed point. 
The system is critical and the correlations functions  
${\cal G} (x,\tau) = \langle {\vec S}(x,\tau) \cdot {\vec S}(0,0)\rangle
 = \langle {\vec T}(x,\tau) \cdot {\vec T}(0,0)\rangle$ display a power law 
behavior.
In the long
distance limit, ${\cal G} (x,\tau)$ is dominated by the contibutions at
$k=0$, $k=2k_F$ and $k=4k_F$:  
\bea
{\cal G}_0 (x,\tau) &\sim& - \frac{3}{4\pi^2} 
\left[ (x + \ri u\tau)^{-2} + (x - \ri u\tau)^{-2}
\right] \nonumber \\
{\cal G}_ {\pi/2}(x,\tau) &\sim& A^2 \cos(\frac{\pi}{2a_0}  x)
(x^2 + u^2\tau^2)^{-3/4} \nonumber \\
{\cal G}_ {\pi}(x,\tau) &\sim& (-1)^{x/a_0} B^2 (x^2 + u^2\tau^2)^{-1},
\label{corb}
\eea
the leading asymptotics thus being ${\cal G}_{\pi/2}$.
In the regions A and C  
the interaction is relevant and leads to the dymanical generation of a mass
gap.
In the far infrared limit, all 
trajectories flow towards the asymptote L: $G = -2 G_3$. 
There the interacting part Hamiltonian (\ref{hpert}) transforms to
\be
{\cal H}_{int} = G/2 \left( \kappa_1 + \kappa_2 + \kappa_6 
- \kappa_3 - \kappa_4 - \kappa_5 \right)^2.
\label{hso6}
\ee
Upon the transformation 
${\vec \xi}_{tR(L)}\rightarrow \pm {\vec \xi}_{tR(L)}$,
the interaction (\ref{hso6}) is easily seen to acquire an SO(6) symmetric
form. However, the conclusion that  
the SO(6) symmetry is restored in both phases A and C would be incorrect.
The scaling portrait in phase C is similar to the crossover sector
of the Kosterlitz-Thouless phase diagram 
for the U(1)-symmetric Thirring model where
an exact (Bethe-ansatz) solution\cite{jap} confirms restoration of SU(2)
up to exponentially small corrections. Using arguments given recently 
by Azaria et al.\cite{azaria1}, we therefore expect that 
restoration of SO(6) is a specific feature of phase C, while
in the massive region A the nature of elementary excitations
is more complicated reflecting the existence of several energy scales.
The development of the strong-coupling regime in the
SO(6) GN model, describing phase C, leads to generation of a fermionic
mass. As a consequence, 
$\langle \kappa_{1,2,6} \rangle = - \langle \kappa_{3,4,5} \rangle \neq 0$,
indicating spontaneous breakdown of translational invariance. Indeed, the
dimerization operators for each chain,
$\Delta_s = (-1)^i {\vec S}_i\cdot{\vec S}_{i+1}$ and  
 $\Delta_t = (-1)^i {\vec T}_i\cdot{\vec T}_{i+1}$,
express in terms of the energy densities of the two SO(3)
spin and orbital GN models:
$\Delta_{s} \sim \kappa_1 + \kappa_2 + \kappa_6 $ and 
$\Delta_{t} \sim \kappa_3 + \kappa_4 + \kappa_5 $.
Therefore
$\langle \Delta_s\rangle = - \langle \Delta_t\rangle 
= \pm \Delta_0$, 
and the system
orders in one of two, doubly degenerate, ground states 
with alternating spin and orbital singlets, in agreement with 
the weak coupling results\cite{nersesyan}.

Calculating the exact dynamical correlation functions in the massive phases
is difficult. While hopeless in the broken-symmetry phase A,
this task could be accomplished in principle in the symmetry-restored phase C
since the SO(6) GN model is integrable.
The full treatment which takes into account the $Z_2$-degeneracy
of the ground state and the existence of topological (kink)
excitations in addition to the fundamental fermion  
will be presented elsewhere. However, since the mass
of the fermion is smaller than twice  the kink mass
we expect that fermions will dominate at sufficiently low energy. 
Their  contribution to the correlation functions 
can be  estimated by a mean field approach:
\bea
\langle {\vec S}(x,\tau) \cdot {\vec S}(y,0)\rangle 
&\sim&   A^2 \cos(\frac{\pi}{2a_0}  x)\cos(\frac{\pi}{2a_0}  y) \; 
K_0(MR) \nonumber \\
 &-& (-1)^{x/a_0} B^2 \; K_0^2(MR) 
 \nonumber \\
\langle {\vec T}(x,\tau) \cdot {\vec T}(y,0)\rangle &\sim& 
A^2 \sin(\frac{\pi}{2a_0}  x)\sin(\frac{\pi}{2a_0}  y) \; 
K_0(MR) \nonumber \\
 &-& (-1)^{x/a_0} B^2 \; K_0^2(MR)
\label{corc}
\eea
where $R= \sqrt{(x-y)^2 + u^2\tau^2}$ and $K_0(MR)$ is the 
real space propagator of a free massive fermion. We observe 
that on top of an {\sl incoherent} background  
at $k \sim \pi$ (with weight $\sim B^2$),
there is  a {\sl coherent} magnon
peak at $k=\pi/2$ (with weight $\sim A^2$). This is to be contrasted with
the situation at weak coupling ($ K \ll J$) where only 
incoherent magnons at $k \sim \pi$
exists\cite{nersesyan}.   
At this point it is worth  commenting on the status of the non-universal 
parameters A and B that enter in the
expressions of the spin densities. The numerical 
results\cite{yamashita,frischmuth} at the SU(4) symmetric point 
are in good agreement with the expressions
(\ref{corb}). In particular, these results have revealed that the peak
in the static susceptibility at $2k_F$ is much greater 
that the one at $4k_F$, thus suggesting 
that $A \gg B$ at the SU(4) point. This has not to be the case when one
deviates from the SU(4) symmetric point, and the question that 
naturaly arises
is how, as $K$ decreases,  will one moves from  strong to weak coupling 
regimes. 
Since our solution for large $K$ captures the properties of both regimes,
it is natural to make the hypothesis that the crossover is encoded in
the $K$ dependence of the nonuniversal constants  A and B. In the simplest 
scenario,
one may conjecture that $B(K)$ will increase as $K$  decreases while $A(K)$
should decrease. Since A is found to be zero at weak coupling, one may further
suspect that it will vanish for $K$ smaller than a critical value $K_D$. Such
a special point where some oscillating component of the correlation function
vanishes is called a disorder point\cite{stephenson}. 

Let us conclude  comparing  our results with the recent DMRG calculations by
Pati et al\cite{pati}. Our result for the phase B is in  agreement
with the numerical data. In the phase C these authors find a doubly degenerate 
ground state which form alternating spin and orbital singlets, in agreement with
our results. However, they conclude that the  mass gap  opens with an exponent 
$ \sim 1.5 \pm 0.25$ whereas the bosonization
approach predicts that the gap is  exponentially small with the deviation from 
the
SU(4) symmetric point.
Moreover, they interpret their datas in favor of $incommensurate$ correlations 
in contrast with (\ref{corc}). In the continuum approach, 
we found no room for incommensuration since the parity breaking (``twist'')
term $ \ri {\vec {\cal N}_a}\left(x+a_0\right) 
\cdot {\vec {\cal N}_a}^{\dagger}\left(x\right) + H. c.$, which appears
upon deviating from the critical SU(4) point and which might be a potential 
source
of incommensutations\cite{nersesyan1}, turns out to be strongly irrelevant
(with dimension 3).
 Our result supports another scenario in which
the correlation functions
contain components at $k \sim 0, \pi$ and $\pi/2 $, with amplitudes 
depending on $K$.

We are grateful to Roland Assaraf, Michel Caffarel, and Michele Fabrizio for illuminating discussions.


\end{document}